\documentclass[11pt]{article}
\usepackage{moriond,epsfig}
\usepackage{psfrag}
\usepackage{xspace}

\def\be{\begin{equation}}
\def\ee{\end{equation}}
\def\bea{\begin{eqnarray}}
\def\eea{\end{eqnarray}}
\def\ra{\rightarrow}
\newcommand{\GeV}{\ensuremath{\mathrm{Ge\kern -0.1em V}}}
\newcommand{\MeV}{\ensuremath{\mathrm{Me\kern -0.1em V}}}
\newcommand{\cm}{\ensuremath{\mathrm{cm}}}
\newcommand{\um}{\ensuremath{\mathrm{\mu m}}}
\newcommand{\bs}{\ensuremath{B_s^0}}
\newcommand{\bd}{\ensuremath{B_d^0}}
\newcommand{\bsd}{\ensuremath{B_{s,d}^0}}
\newcommand{\bu}{\ensuremath{B^{+}}}
\newcommand{\mm}{\ensuremath{\mu^{+}\mu^{-}}}

\newcommand{\bsmm}{\ensuremath{\bs\ra\mm}}
\newcommand{\bdmm}{\ensuremath{\bd\ra\mm}}
\newcommand{\bsdmm}{\ensuremath{\bsd\ra\mm}}
\newcommand{\bjk}{\ensuremath{\bu\ra J/\psi K^{+}}}

\newcommand{\bjf}{\ensuremath{\bs\ra J/\psi \phi}}
\newcommand{\bpf}{\ensuremath{\bs\ra \psi(2S) \phi}}
\newcommand{\brbsmm}{\ensuremath{\mathcal{B}(\bsmm)}}
\newcommand{\brbdmm}{\ensuremath{\mathcal{B}(\bdmm)}}
\newcommand{\brbsdmm}{\ensuremath{\mathcal{B}({B_{s,d}^0}\ra\mm)}}

\newcommand{\bsphimumu}{\ensuremath{\bs \ra \phi \,\mu^+ \mu^-}}

\newcommand{\Mmm}{\ensuremath{M_{\mu\mu}}}

\newcommand{\Lxyz}{\ensuremath{{L}_{\rm{3D}}}}
\newcommand{\ctau}{\ensuremath{\lambda}}
\newcommand{\pting}{\ensuremath{\Delta\Theta}}
\newcommand{\iso}{\ensuremath{\mathit{I}}}
\newcommand{\ptmm}{\ensuremath{\vec{p}^{\:\mu\mu}_{T}}}
\newcommand{\pmm}{\ensuremath{\vec{p}^{\:\mu\mu}}}

\newcommand{\ppbar}{$p\overline{p}$\xspace}
\newcommand{\CME}       {center-of-mass energy\xspace}
\newcommand{\tev}{\mbox{\,\rm Te\kern -0.1emV}}
\newcommand{\gev}{\mbox{\,\rm Ge\kern -0.1emV}}
\newcommand{\mev}{\mbox{\,\rm Me\kern -0.1emV}}
\newcommand{\kev}{\mbox{\,\rm ke\kern -0.1emV}}
\newcommand{\fb}        {\xspace{fb$^{-1}$}\xspace}
\newcommand{\pb}        {\xspace{pb$^{-1}$}\xspace}
\newcommand{\dzero}{D\O \xspace}

\begin{document}
\vspace*{4cm}
\title{Search for rare decays of the $B_s$ Meson at the Tevatron}

\author{Ralf Patrick Bernhard}

\address{Institute of Physics, Winterthurerstrasse 190,\\
8057 Z\"urich, Switzerland}

\maketitle

\vskip -5 cm
\centerline{\includegraphics[width = 4.4 cm]{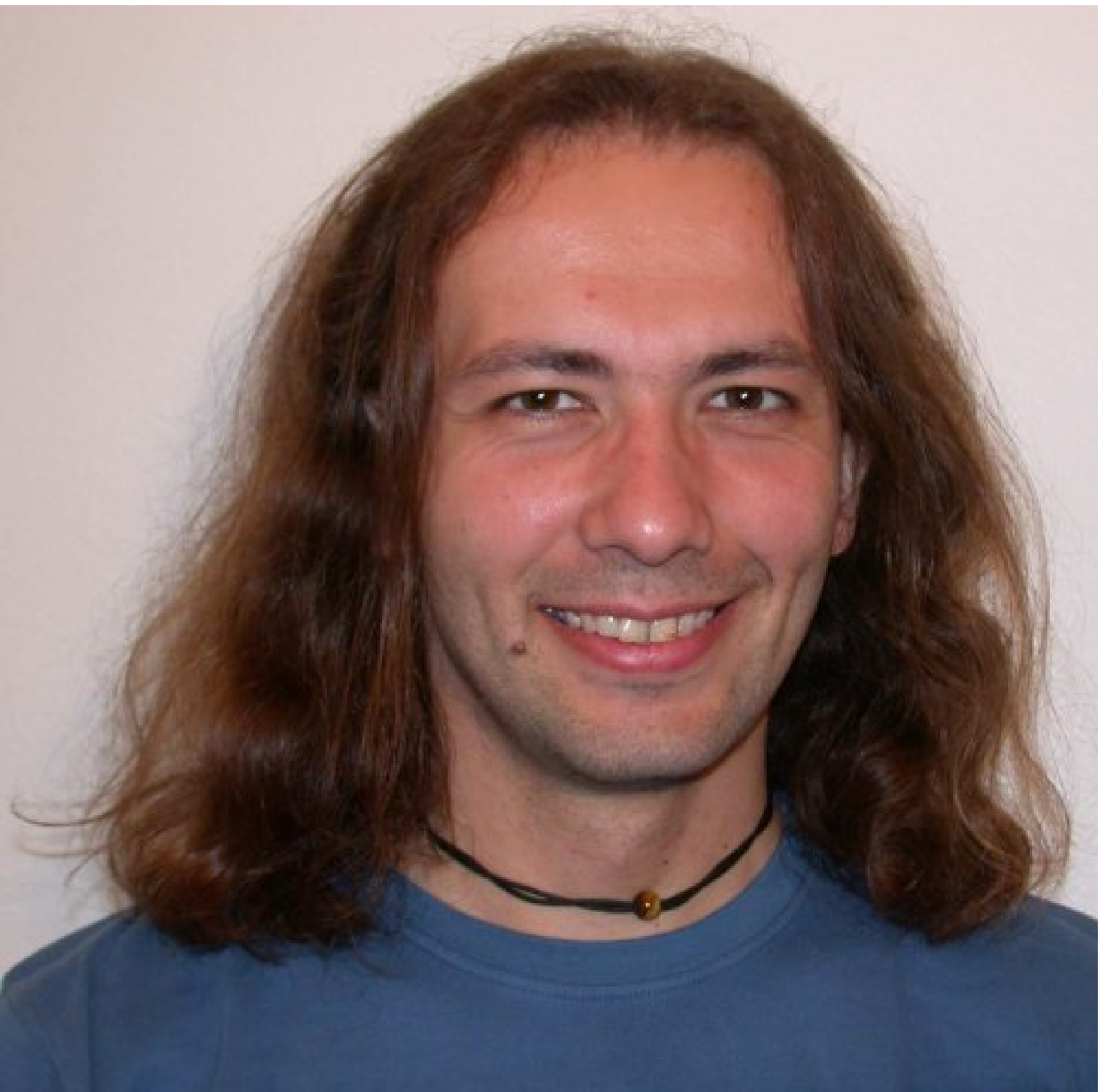} }
\vskip 0.5 cm
\abstracts{
We report on searches for Flavor-Changing Neutral Current (FCNC) decays in \ppbar collisions at a \CME of 1.96 \tev~using up to 0.78~\fb of data, collected at the CDF and D\O~detectors respectively. The rare FCNC decays presented here are the searches for \bsdmm\ and \bsphimumu.
}

\section{Introduction}
In the Standard Model (SM), Flavor-Changing Neutral Current (FCNC) decays are
highly suppressed and can only occur through higher order diagrams.
For example the  branching fractions of the FCNC decays \bsdmm\ have a 
SM expectation~\cite{smbr} of
$\brbsmm=(3.42\pm0.54)\times10^{-9}$ and
$\brbdmm=(1.00\pm0.14)\times10^{-10}$. These predictions are about two orders
of magnitude smaller than the current experimental sensitivity.
However, new physics contributions can significantly enhance these branching
fractions. In the absence of an observation, any improvements 
to the limits can be used to set significant constraints on various
models beyond the SM.
The exclusive FCNC decay \bsphimumu\ is related to the quark-level transition of $b \rightarrow s\,
\ell^+\ell^-$. The long term goal is  study kinematic properties for these decay like the invariant di-muon mass distribution or the forward backward asymmetry.
But also an observation of this decay or experimental upper limit on its rate will yield additional important information on the flavor dynamics of FCNC decays.

\section{Search for the rare decays \bsdmm}

The most stringent published limits are a result of a combination~\cite{BsMMCombo} of the recent \dzero\ result~\cite{D0BsMMprl} using 0.3 \fb of data and the latest published limit from CDF~\cite{CDFBsMMprl} using 0.36 \fb of recorded data. 
The obtained limits  at 95\% C.L. were ${\cal B}(\bsmm) < 1.5 \times 10^{-7}$ and ${\cal B}(\bdmm) < 4.0 \times 10^{-8}$ respectively.

The used analysis procedure is similar in the \dzero and  CDF experiments.    
After a pre-selection step, special discriminating variables in an optimization procedure are used to further enhance the signal efficiency while reducing the expected background. Both collaborations employ a ``blind'' analysis strategy when choosing the
final selection criteria. As normalization the well measured and high statistics decay \bjk\ was used. 
In the following the most recent analyses from both experiments will be described. 

The strategy of the \dzero~collaboration is to use the already published analysis procedure~\cite{D0BsMMprl} for the data set recorded after  the first publication, to obtain a sensitivity. This obtained sensitivity is then combined with the already obtained limit. CDF used a published analysis procedure~\cite{CDFBsMMprl} as well and added more data to get an updated limit.
 
\dzero~used data collected by dimuon tiggers with two muons of opposite charge that form a
common secondary 3D-vertex with an invariant mass between 4.5 and 7.0 GeV/$c^2$. Each muon candidate had to have $p_T > 2.5~$GeV/$c$, $|\eta| < 2.0$ and a sufficient number of hits in the central
tracking station. To ensure a similar $p_T$ dependence of the $\mu^+\mu^-$ system in the signal and in
the normalization channel, $p_T^B$ had to be greater than 5~GeV/$c$.

\begin{figure}[h]
\begin{minipage}[c]{0.49\textwidth}
\begin{center}
  \includegraphics[width=\linewidth]{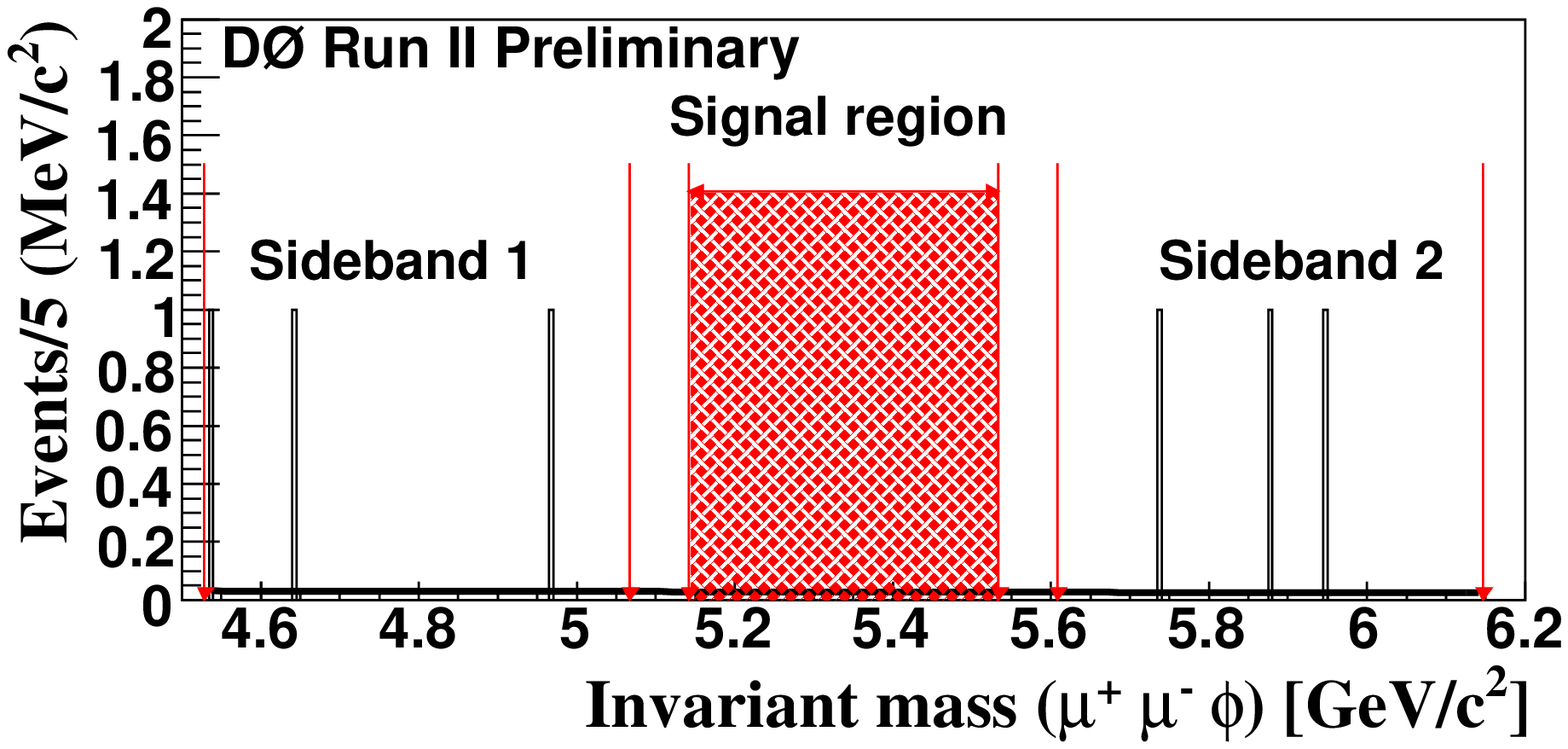}
\end{center}
\end{minipage}
\begin{minipage}[c]{0.49\textwidth}
\begin{center}
   \includegraphics[width=\linewidth]{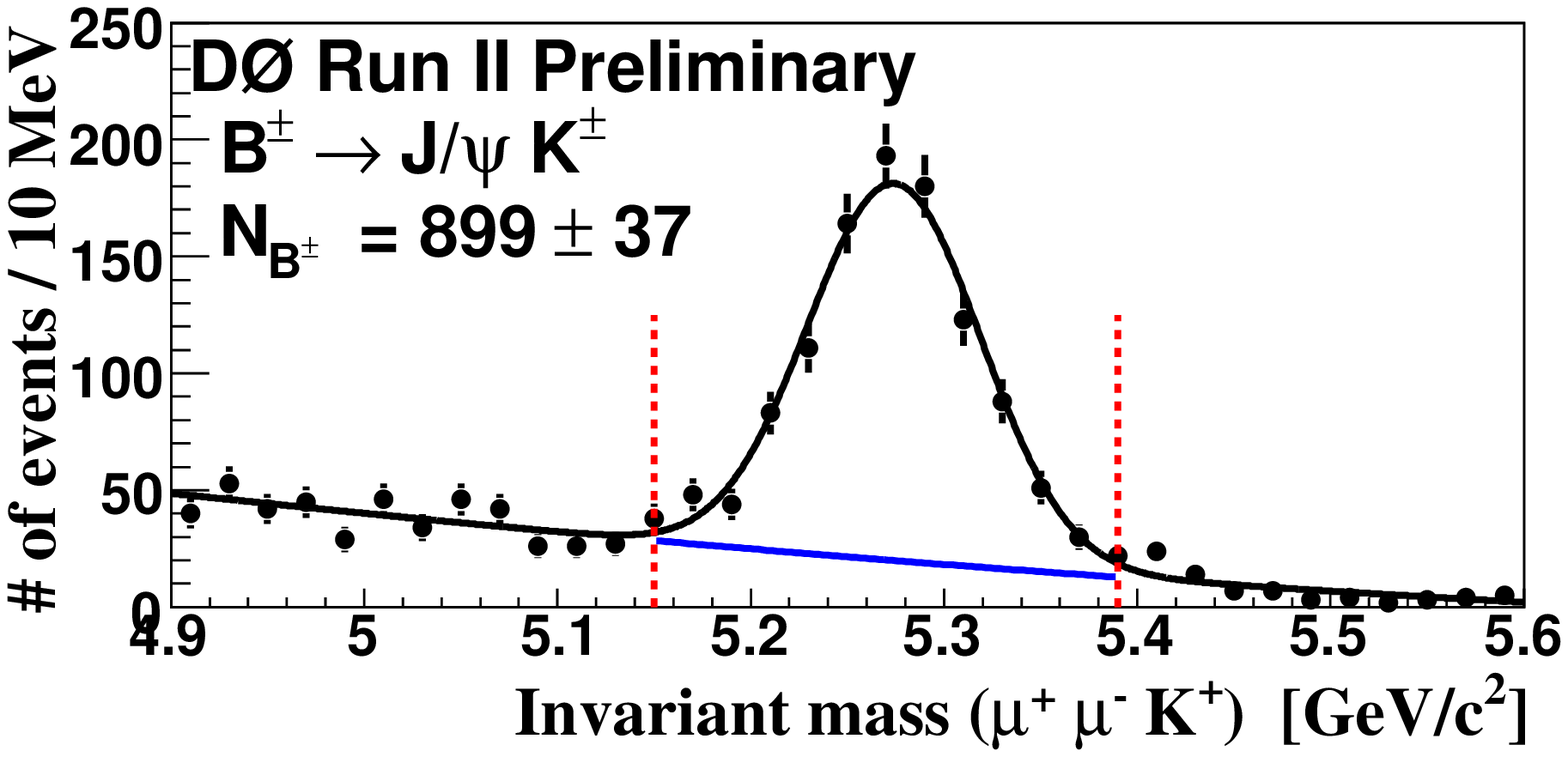}
\end{center}
\end{minipage}
  \caption{\label{fig_background_box_closed}The invariant mass
    distribution for the decay $\bsmm$ after optimized requirements on the discriminating variables on the left side. On the right side the invariant mass distribution of the normalization channel  \bjk\  is shown.}
\end{figure}

For the \dzero analysis the following variables were found to exploit best the distinctive properties of the $B$ meson: the 3D Opening-Angle, \pting\ (here the decay
length vector is determined in three-dimensions), the $B$-candidate Isolation,
\iso, and the Transverse Decay Length Significance, $L_{2D}/\sigma_{L_{2D}}$.
After optimizing the cuts and a linear interpolation of the sideband population for the data sample into the signal region we obtain an expected number of background events of 2.2$\pm$0.7. The remaining background distribution is shown
in Fig.~\ref{fig_background_box_closed} (left side). 
The values of the discriminating variables for this additional data set changed only slightly with respect to the previous analysis. The resulting sensitivity is calculated using the \bjk\  decay as normalization. The invariant mass distribution of the decay channel is shown in Fig.~\ref{fig_background_box_closed} (right side).
This yields a sensitivity including all the statistical and systematical uncertainties of
$\langle {\cal B}(B_s^0 \rightarrow \mu^+ \mu^-)\rangle < 3.0\,(3.7)\times 10^{-7}$
at the 90\% (95\%) C.L.

The obtained expected limit for the new data set is combined with the previous analysis~\cite{D0BsMMprl} using 0.3 \fb of data. 
The expected limit for the entire data set is thus a combination of two measurements: one actual measurement with observed events (old data) and one hypothetical experiment (new data) with all possible outcomes in the signal region weighted by their Poisson probability of occurrence, i.e., under the assumption of a background hypothesis only. We have used a Bayesian technique to combine the two experiments and include at this stage the uncorrelated uncertainties into the expected limit calculation. The background uncertainty, the uncertainty on the number of normalization events and the statistical error on the efficiencies are considered as uncorrelated.
The expected upper limit at the 90\% (95\%) C.L. for the entire D\O \ data set of 0.7 \fb is 
\begin{eqnarray*}
\langle {\cal B}(B_s^0 \rightarrow \mu^+ \mu^-)\rangle < 1.9\,(2.3)\times 10^{-7}.
\end{eqnarray*}

The CDF detector has of two muon detectors with different acceptances.
Four layers of planar drift chambers (CMU) detect muon candidates with 
$p_T>1.4~\GeV/c$ and provide coverage in the pseudorapidity range 
$\left|\eta\right|<0.6$. The central muon extension (CMX) consists of conical 
sections of drift tubes and extends the coverage to $0.6<\left|\eta\right|<1.0$ 
for muon candidates with $p_T>2.0~\GeV/c$.  
The data used in the analysis are selected by two classes of dimuon 
triggers: for the CMU-CMU (U-U) triggers both muon candidates are reconstructed 
in the CMU chambers, while for the CMU-CMX (U-X) triggers one of the muon 
candidates is reconstructed in the CMX chambers.  
Since they have different sensitivities, the U-U and U-X channels are treated separately, 
combining the results at the end.
The offline reconstruction begins by identifying two muon candidates of 
opposite charge which satisfy the online dimuon trigger requirements and  
with $p_{T}> 2\ (2.2) \;\GeV/c$.  
The random combinatoric backgrounds are suppressed by requiring the vector sum 
of the muon transverse momenta to be $| \ptmm | > 4\;\GeV/c$.  
The 3D decay length is given by $\Lxyz=\vec{L}\cdot\pmm/|\pmm|$, 
where $\vec{L}$ is the displacement vector from the primary to the dimuon vertex.  
The primary vertex is determined using a constrained vertex fit of all  
tracks in the event, excluding the \mm\ pair and other secondary decay tracks.  
For each $B$-candidate the proper decay time $\tau=\Mmm\Lxyz/|\pmm|$ is estimated, where
\Mmm\ is the invariant mass and \pmm\ is the momentum vector of the dimuon system.
Additional background is reduced by demanding $\Lxyz < 1.0~\cm$, the 
uncertainty on $\Lxyz$ to be less than 150 \um, and  
$2\sigma_{\ctau} < \ctau <0.3~\cm$, where $\ctau=c\cdot \tau$ and $\sigma_{\ctau}$ 
is the total uncertainty on $\ctau$.  

\begin{figure}[h]
\begin{minipage}[c]{0.49\textwidth}
\begin{center}
  \includegraphics[width=\linewidth]{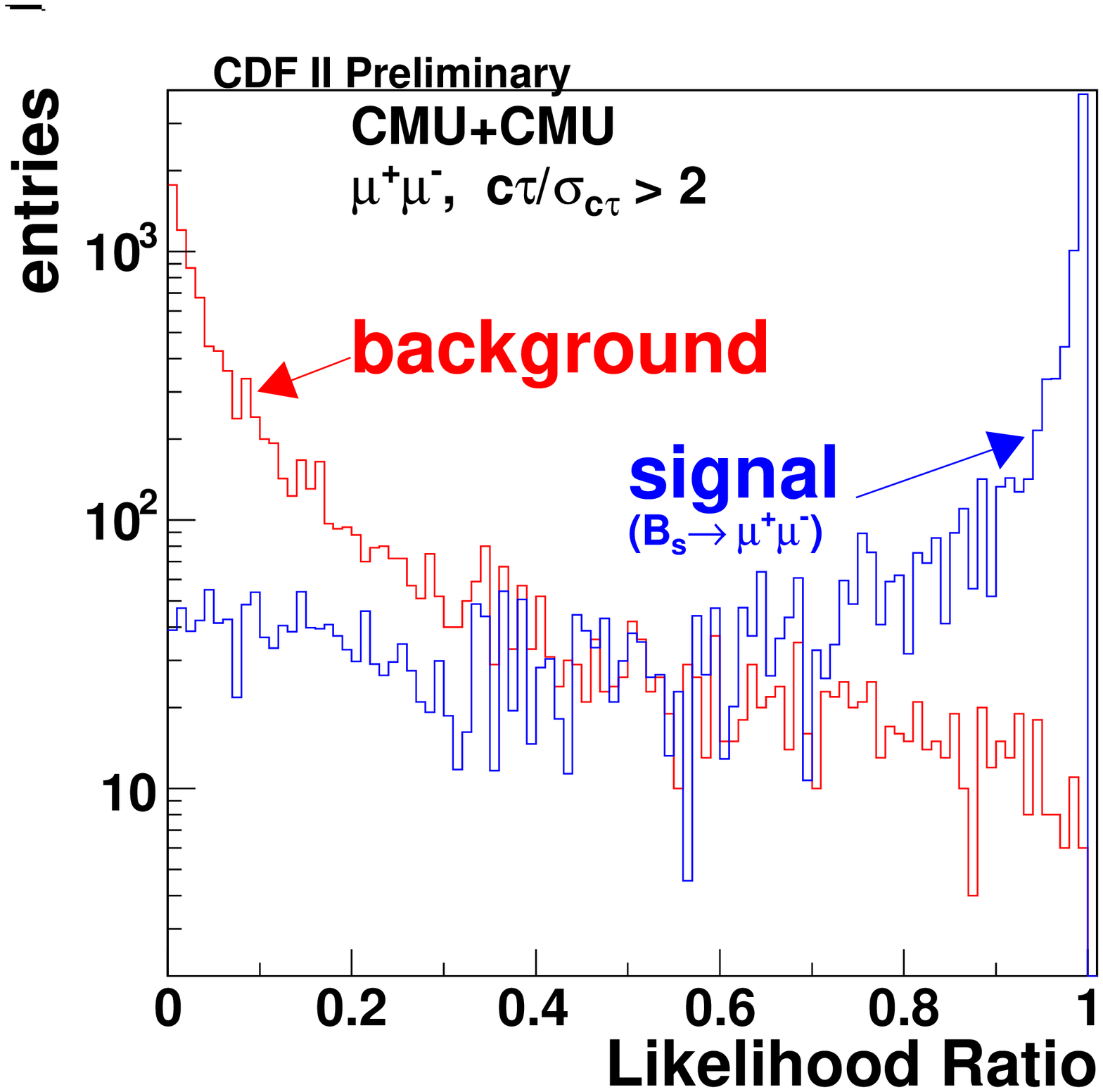}
\end{center}
\end{minipage}
\begin{minipage}[c]{0.49\textwidth}
\begin{center}
   \includegraphics[width=\linewidth]{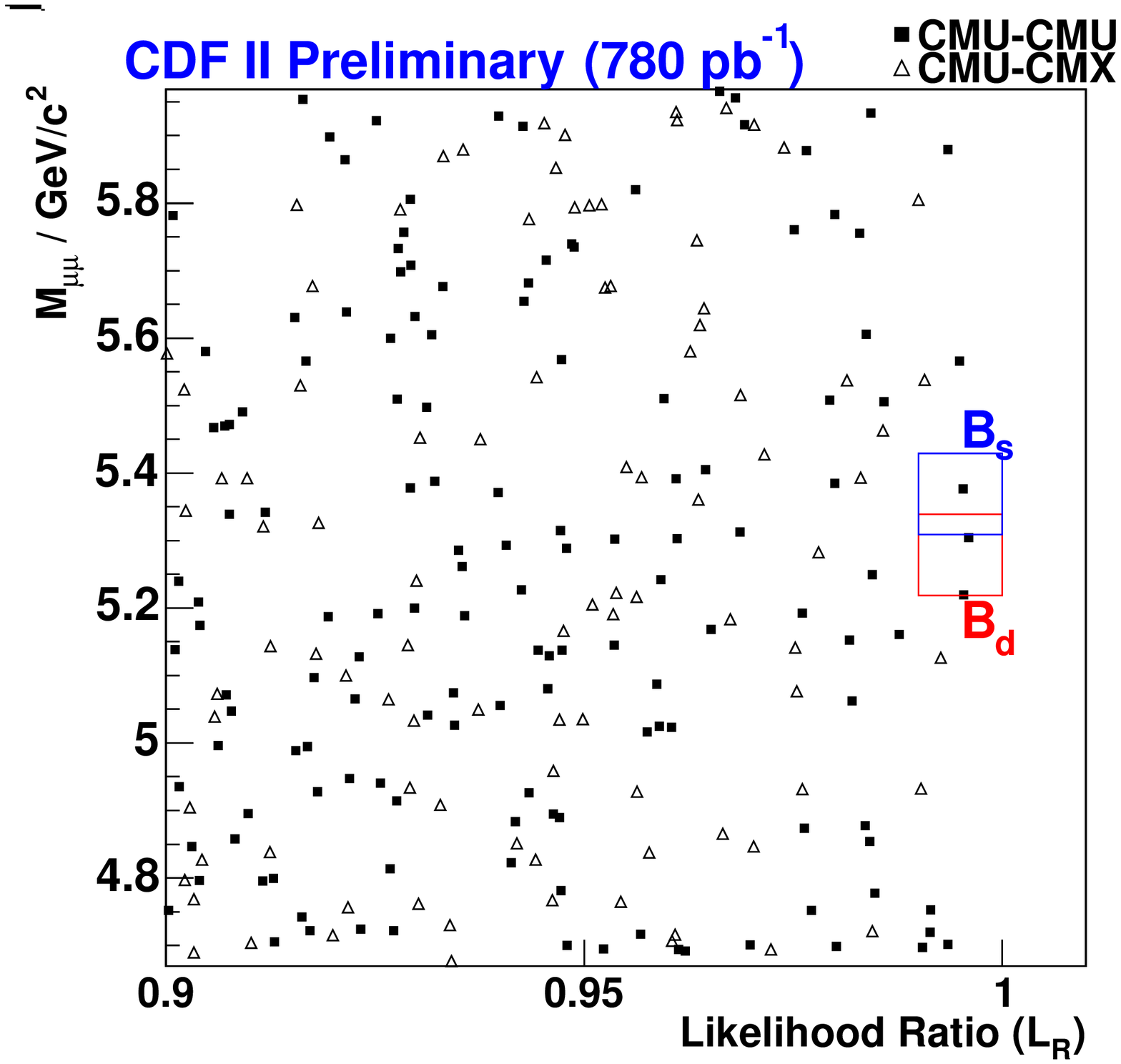}
\end{center}
\end{minipage}
  \caption{\label{fig:CDFBsMMResults} Signal and background distribution of the likelihood ratio for the U-U channel on the left side. Invariant mass distribution versus the event likelihood with the signal boxes indicated on the right side.}
\end{figure}

To enhance signal and background separation CDF constructs a multivariate
likelihood ratio based on the input variables: \iso, \pting, and
\ctau\ probability $P(\ctau)=e^{-\ctau/c\tau_{B_{s(d)}}}$, where
$\tau_{B_{s(d)}}$ is the world average $B_{s(d)}$ lifetime.  CDF uses the
$P(\ctau)$ variable instead of $\ctau$ in constructing the likelihood
ratio because the $P(\ctau)$ distribution is nearly flat, and therefore
better behaved in the likelihood.  The likelihood ratio is then defined to be
\begin{equation}\label{eq:LH}
  L_R = \frac{\prod_{i} \mathbf{P}_{s}(x_{i})}
       {\prod_{i} \mathbf{P}_{s}(x_{i})
      + \prod_{i} \mathbf{P}_{b}(x_{i})},
\end{equation}
where $x_1=\iso$, $x_2=\pting$, $x_3=P(\ctau)$, and $\mathbf{P}_{s(b)}(x_{i})$
is the probability that a signal (background) event has an observed $x_i$.
The probability distributions for the signal events are obtained from the signal
MC and the background distributions are taken from the data
sidebands. An example of the likelihood ratio for the U-U channel is shown in Fig~\ref{fig:CDFBsMMResults} (left side). 
The optimization of the analysis is based on the {\it{a priori}} expected $90\%$~C.L. upper limit on \brbsdmm.  The expected limit for a given set of optimization requirements is computed by summing the $90\%$~C.L. limits over all possible observations $N_o$, weighted by the corresponding Poisson probability when expecting $N_b$ background events.
A scan over a range of $L_R$ requirements
the optimal value was determined to be $L_R>0.99$.  With the optimized selection requirements, 
the expected number of background events in 0.78 \fb for \bs\ in a signal region of $\pm 60~\MeV/c^2$ 
($\pm 2.5\sigma_{M}$) around the world average \bs\ mass~\cite{pdg} is
$N_{b}$ is $0.88\pm0.30$ [$0.39\pm0.21$]  for the U-U [U-X] channel. The expected number of background events in a $\pm 60~\MeV/c^2$  
signal region around the \bd\  is estimated to be $N_{b}$ is $1.86\pm0.34$ [$0.59\pm0.21$]  in the U-U [U-X] channel.
Using these criteria 1 [0] and 2 [0] events are observed in the U-U [U-X] channel of the  \bs\ and \bd\ signal boxes respectively which is consistent
with the background expectations. 
shown in Fig.~\ref{fig:CDFBsMMResults} (right side). 
Combining the U-U and U-X channels taking into
account the correlated uncertainties, 90\% (95\%) C.L. limits can be derived to be
\begin{eqnarray*}
\brbsmm &<& 8\times 10^{-8} (1.0\times 10^{-7}) \, \, \, {\rm and} \\
\brbdmm &<& 2.3\times 10^{-8} (3.0\times 10^{-8}).  
\end{eqnarray*}

\section{Search for the rare decays \bsphimumu}
The decay \bsphimumu\ is an exclusive FCNC in the \bs\ meson system. 
Within the SM, the decay rate for the \bsphimumu\ decay, neglecting the interference effects with the much stronger \bjf\ and \bpf\ resonance decays, is predicted to be of the order of $1.6\times 10^{-6}$~(see Ref.~\cite{geng}) with about 30\% uncertainty due to poorly known form factors. 

\begin{figure}[h]
\begin{minipage}[c]{0.49\textwidth}
\begin{center}
  \includegraphics[width=\linewidth]{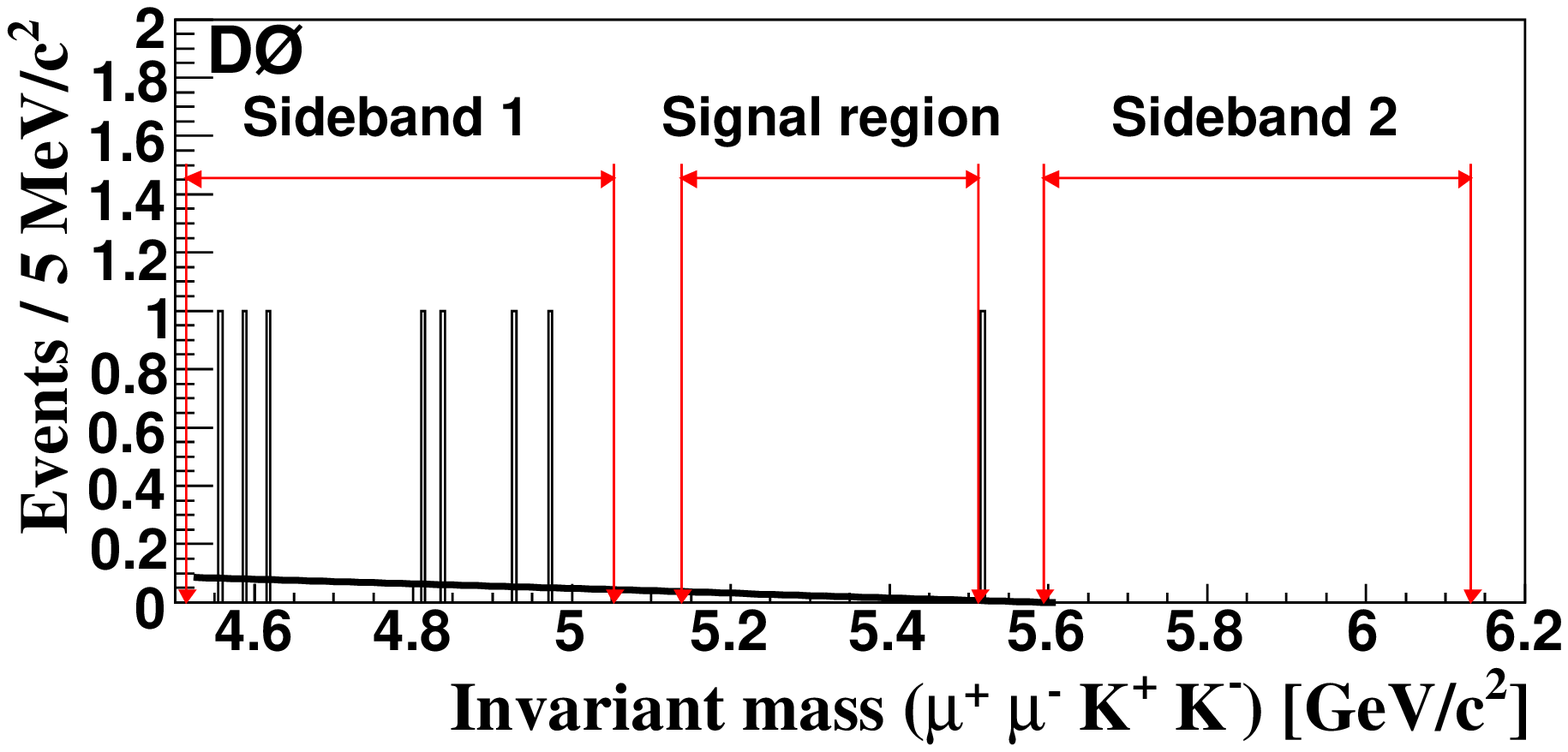}
\end{center}
\end{minipage}
\begin{minipage}[c]{0.49\textwidth}
\begin{center}
\includegraphics[width=\linewidth]{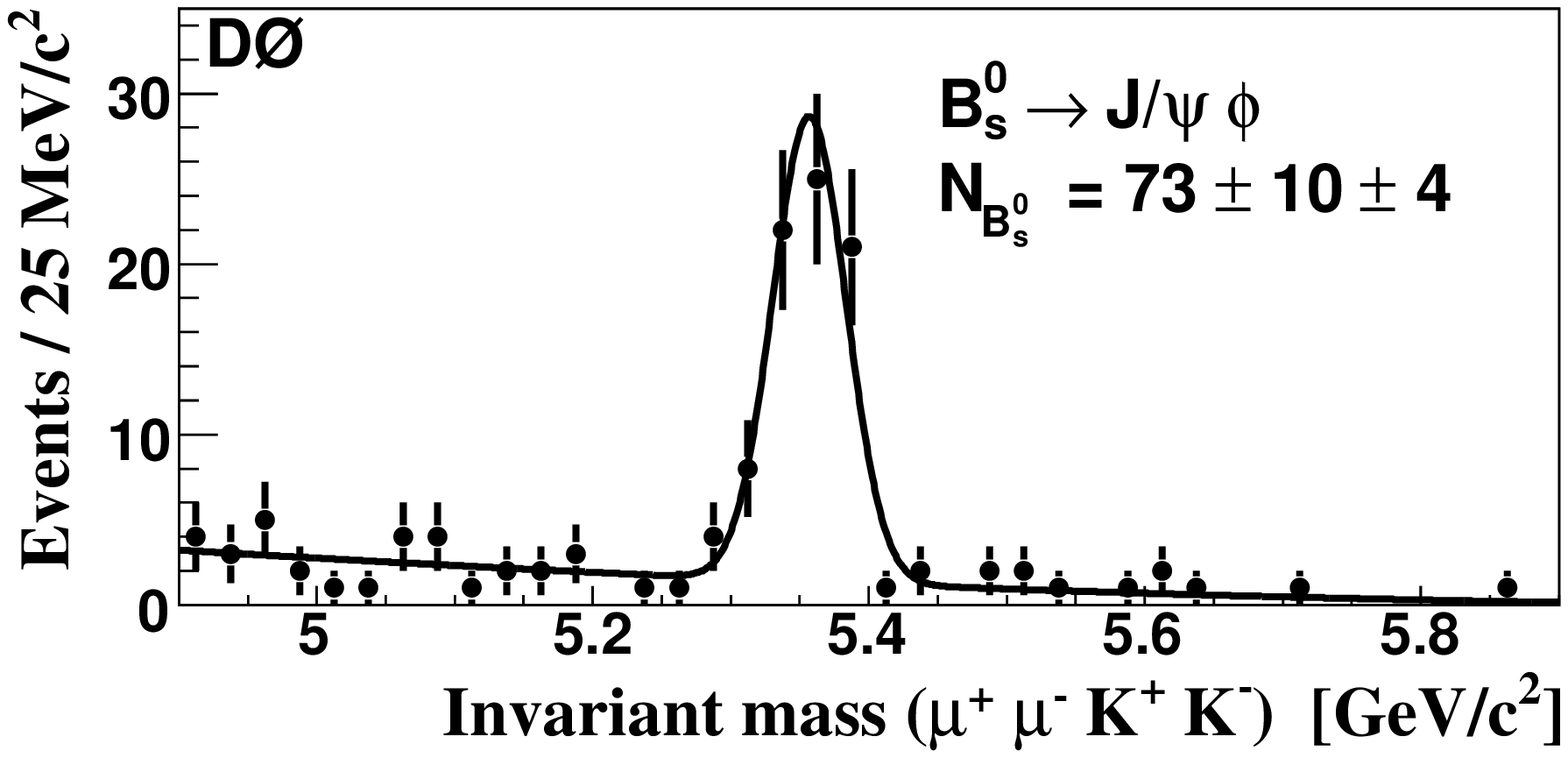}
\end{center}
\end{minipage}
  \caption{\label{fig_background_bsphimumu}The invariant mass
    distribution for the decay $\bsphimumu$ on the left side.
	On the right side the invariant mass distribution of the normalization channel $\bjf$ is shown.
}
\end{figure}

A search for this rare decay is performed by the \dzero collaboration and uses about 0.45~\fb~of data. The event pre-selection starts by requiring exactly two muons fulfilling quality cuts on the number of hits in the muon system and the two additional charged particle tracks to form a good vertex. The invariant mass of the two muons is required to be within
$0.5<m_{\mu^+\mu^-}<4.4$~GeV/$c^2$. In this mass region, the $J/\psi(\rightarrow \mu^+\mu^-)$ and $\psi(2S)(\rightarrow \mu^+\mu^-)$ resonances are excluded to discriminate against dominant resonant decays. 
The $\chi^2/d.o.f.$ of the two-muon vertex is required to be less than
10.  
The tracks that are matched to each muon are required to have
a sufficient number of hits in the tracking detectors 
and the transverse momentum of each of the muons is required to be greater
than 2.5~GeV/$c$ with $|\eta|<2.0$ to be well inside the fiducial
tracking and muon detector acceptances.
The number of $\bsphimumu$ candidates is further
reduced by requiring $p_T^B>5~$GeV/$c$ and asking the $\bs$
candidate vertex to have $\chi^2 < 36$ with 5 $d.o.f.$ The two
tracks forming the $\phi$ candidate are further required to have
$p_T>0.7$~GeV/$c$ and their invariant mass within the range $1.008
<m_{\phi} < 1.032~$GeV/$c^2$. In analogy to the search for the decay \bsmm the same discriminating variables and optimization procedure is used to enhance the sensitivity. The optimal values of the discriminating variables yield after a linear interpolation of the sidebands into the mass window signal region, 
1.6$\pm$0.4 expected background events as shown in Fig.~\ref{fig_background_bsphimumu} (left side). Upon examining the data in the mass region, zero candidate events are observed in the signal region, consistent with the background events as
estimated from sidebands. The Poisson probability of observing zero events for an expected background of $1.6\pm0.4$ is $p=0.22$.
In the absence of an apparent signal, a limit on the branching fraction ${\cal B}(\bsphimumu)$ can be computed by normalizing the upper limit on
the number of events in the $B^0_s$ signal region to the number of reconstructed $\bjf$ events.
The mass spectrum of the reconstructed $\bjf$ after all cuts is shown in Fig.~\ref{fig_background_bsphimumu} (right side).
Including statistical and systematic uncertainties, without including the uncertainty on the measured branching fraction of the decay $\bjf$, the Feldman and Cousins (FC) limit is
\begin{eqnarray*}
\frac{{\cal B}(\bsphimumu)}{{\cal B}(\bjf)} <4.4\,(3.5)\,\times 10^{-3}
\end{eqnarray*}

at the 95\% (90\%) C.L. respectively. Taking a Bayesian approach with a flat prior and the
uncertainties treated as Gaussian distributions in the integration, an
upper limit of ${\cal B}(\bsphimumu)/{\cal B}(\bjf) <7.4 \,(5.6)
\times 10^{-3}$ at the 95\% (90\%) C.L., respectively was found.
Using only the central value of the world average branching fraction~\cite{pdg} of
${\cal B}(\bjf) = (9.3 \pm 3.3)\,\times 10^{-4}$, the FC limit
corresponds to ${\cal B}(\bsphimumu) < 4.1\,(3.2)\,\times 10^{-6}$ at
the 95\% (90\%) C.L. respectively. This is presently the most stringent upper bound and has been recently submitted to PRL~\cite{Abazov:2006qm}.

\end{document}